\documentclass[aps,amssymb,pra,showpacs,twocolumn]{revtex4}

\usepackage{epsf}
\usepackage{amsmath}
\usepackage{graphicx}

\begin{document}

\title{Two-dimensional tunneling in a SQUID}

\author{B. Ivlev} 

\affiliation
{Instituto de F\'{\i}sica, Universidad Aut\'onoma de San Luis Potos\'{\i}\\
San Luis Potos\'{\i}, S. L. P. 78000 Mexico}


\begin{abstract}
Traditionally quantum tunneling in a static SQUID is studied on the basis of a classical trajectory in imaginary time under a two-dimensional potential barrier. The trajectory connects
a potential well and an outer region crossing their borders in perpendicular directions. In contrast to that main-path mechanism, a wide set of trajectories with components tangent to the border of the well can constitute an alternative mechanism of multi-path tunneling. The phenomenon is essentially non-one-dimensional. Continuously distributed paths under the barrier result in enhancement of tunneling probability. A type of tunneling mechanism (main-path or multi-path) depends on character of a state in the potential well prior to tunneling.
\end{abstract} \vskip 1.0cm

\pacs{85.25.Dq, 74.50.+r}

\maketitle

\section{INTRODUCTION}
\label{intr}
Despite phases in Josephson junctions \cite{JOS} are treated as macroscopic degrees of freedom they can exhibit quantum properties \cite{LEGGETT}. In particular, quantum tunneling of those variables across a potential barrier is possible \cite{LEGGETT1,OVC1,CLARKE1,CLARKE2}. Tunneling in a single Josephson junction is similar to a conventional one-dimensional quantum mechanical process. In this case the tunneling mechanism is described by theory of Wentzel, Kramers, and Brillouin (WKB) \cite{LANDAU}. Tunneling occurs from a classically allowed region which is a conventional potential well where energy levels are quantized \cite{CLARKE3,CLARKE4,UST1,UST2}. Quantum coherence between potential wells was demonstrated \cite{LEGGETT2,NAK,TOLP,MOO}.

Besides single Josephson junctions superconducting quantum interference devices (SQUID) are also a matter of active investigation for many years
\cite{CHEN,OVC2,CRIST,WANG,BAL,CAST,MITR,BLAT}. A SQUID consists of two Josephson junctions and, therefore, represents a two-dimensional system where macroscopic quantum tunneling is also possible. Tunneling in multi-dimensional systems is well studied \cite{COLEMAN1,COLEMAN2,SCHMID1,SCHMID2,MELN}. There is the certain underbarrier path (or a few paths) where a wave function is localized and it decays along the path. This main-path tunneling is described by a classical trajectory in imaginary time and it is generic with a conventional WKB mechanism.

In contrast to that, a different scenario of tunneling in a static SQUID is possible. Instead of localization on a main path an underbarrier state is distributed over a continuous
set of paths. Multi-path tunneling cannot be described in terms of a classical trajectory in imaginary time.

A realization of a particular tunneling mechanism (main-path or multi-path) depends on type of a state in the potential well prior to tunneling. Suppose, in the classical language, a state in a well to have a momentum component orthogonal at some point to a border of the well (a normal reflection). The main path starts at that point and continues under the barrier. When a state in the well has also a component which is tangent to the border at some point (a particle hits the border from some angle) a scenario is different. Around that point there a set of continuously distributed classical paths which go under the barrier.

A goal of the paper is to propose the different aspect of tunneling in a static SQUID which is multi-path mechanism. The phenomenon is essentially non-one-dimensional. As shown in the paper, a multi-path mechanism can result in a larger probability of tunneling compared to main-path. Different paths interfere and a method of classical trajectory in imaginary time, in contrast to main-path, is not valid. In experiments it is possible to determine a contribution of multi-path effects to a total probability of tunneling. A symmetric dc SQUID in zero magnetic field and without dissipation is considered below.

In Sec.~\ref{formulation} a formulation of the problem is given. In Sec.~\ref{gen} general arguments are used to explain the phenomenon and to estimate the effect. It is shown that multi-path tunneling through a two-dimensional static barrier reminds photon-assisted tunneling across a nonstationary one-dimensional barrier. In Sec.~\ref{ham-jac} an exact solution of the semiclassical problem is done with the use of a certain model coupling between the junctions in a SQUID. The exact calculations confirm the estimate of the effect performed on the basis of general arguments in Sec.~\ref{gen}.
\section{FORMULATION OF THE PROBLEM}
\label{formulation}
We consider a dc SQUID, consisting of two identical Josephson junctions with phases $\varphi_{1}$ and $\varphi_{2}$, with no dissipation when the two junctions are inductively coupled. A classical behavior of phases corresponds to conservation of the total energy
\begin{eqnarray}
\label{1}
\nonumber
&&E_{0}=\frac{E_{J}}{2\omega^{2}}\left[\left(\frac{\partial\varphi_{1}}{\partial t}\right)^{2}+
\left(\frac{\partial\varphi_{2}}{\partial t}\right)^{2}\right]+E_{J}\bigg[-\cos\varphi_{1}\\
&&-\cos\varphi_{2}-\frac{I}{2I_{c}}(\varphi_{1}+\varphi_{2})+\frac{1}{2\beta}(\varphi_{1}-\varphi_{2})^{2}\bigg],
\end{eqnarray}
where the Josephson energy $E_{J}=\hbar I_{c}/2e$, the plasma frequency $\omega=\sqrt{2eI_{c}/\hbar C}$, and the coupling parameter $\beta=2\pi LI_{c}/\Phi_{0}$ are introduced. Here $I_{c}$, $L$, and $C$ are critical current, inductance, and capacitance of each individual junction. The magnetic flux quantum is $\Phi_{0}=\pi\hbar c/e$.
\begin{figure}
\includegraphics[width=5cm]{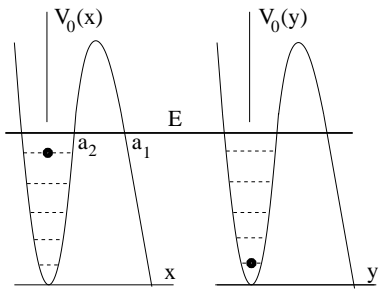}
\caption{\label{fig1}Uncoupled junctions, $\alpha=0$. The total particle energy $E$ is fixed. The maximal 
tunneling probability in the $x$ direction relates to a minimal excitation of the $y$ motion to provide a maximal energy in
the $x$ direction.}
\end{figure}

Below we consider large $\beta$ and the total current $I$ close to its critical value
\begin{equation}
\label{2}
\left(1-\frac{I}{2I_{c}}\right)\ll 1
\end{equation}
and introduce new variables by the relations
\begin{eqnarray}
\label{3}
&&\varphi_{1}=\frac{\pi}{2}+(3x-1)\sqrt{2\left(1-I/2I_{c}\right)}+\frac{3x}{\beta}\\
\nonumber
&&\varphi_{2}=\frac{\pi}{2}+(3y-1)\sqrt{2\left(1-I/2I_{c}\right)}+\frac{3y}{\beta}.
\end{eqnarray}
Below time is measured in the unit of $t_{0}$ defined by
\begin{equation}
\label{3a}
\frac{1}{t_{0}}=\frac{\omega}{\sqrt{2}}\left[\sqrt{2\left(1-I/2I_{c}\right)}+1/\beta\right]^{1/2}.
\end{equation}
The energy (\ref{1}) takes the form
\begin{equation}
\label{4}
E_{0}=\frac{\hbar B}{t_{0}}\left[\frac{1}{2}\left(\frac{\partial x}{\partial t}\right)^{2}+
\frac{1}{2}\left(\frac{\partial y}{\partial t}\right)^{2}+V(x,y)\right],
\end{equation}
where
\begin{equation}
\label{5}
B=\frac{9E_{J}}{\hbar\omega\sqrt{2}}\left[\sqrt{2\left(1-I/2I_{c}\right)}+1/\beta\right]^{5/2}.
\end{equation}
The potential energy is
\begin{equation}
\label{6}
V(x,y)=V_{0}(x)+V_{0}(y)-\frac{2\alpha xy}{1+\alpha},
\end{equation}
where $V_{0}(x)=x^{2}-x^{3}$ and the coupling parameter is
\begin{equation}
\label{7}
\alpha=\frac{1}{\beta\sqrt{2(1-I/2I_{c})}}.
\end{equation}
$B$ in Eq.~(\ref{5}) is called semiclassical parameter. When $B$ is large the phase dynamics is mainly classical. Below we consider that case, $1\ll B$.
\begin{figure}
\includegraphics[width=5.5cm]{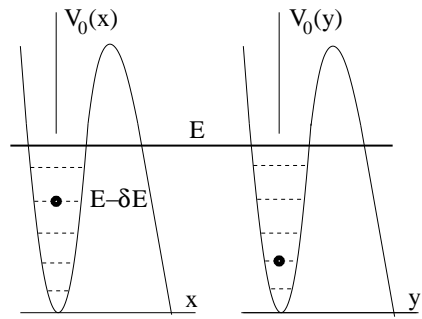}
\caption{\label{fig2}Uncoupled junctions, $\alpha=0$. A part of the total energy $E$ is given up to an excitation of the $y$ motion resulting in a reduced tunneling probability in the $x$ direction compared to Fig.~\ref{fig1}.}
\end{figure}
\section{TUNNELING IN TWO DIMENSIONS}
\label{gen}
A classical dynamics of phases in a SQUID is described by Eqs.~(\ref{4}) and (\ref{6}). The effective particle moves in the classically allowed region, in a vicinity of the point $x=y=0$, which is restricted by the potential barrier. As known, the particle can tunnel through the barrier resulting in experimentally observable phase jumps. Character of tunneling
depends on coupling strength $\alpha$ between the two junctions. When the coupling is strong, $1\ll\alpha$, one can easily show that the last term in the energy (\ref{1}) dominates
\cite{OVC2}. Therefore in this case the particle tunnels along the direction $x=y$ since fluctuations around this path cost a large energy. In contrast, at a small coupling, $\alpha\ll 1$, the junctions are almost independent and there are two different tunneling paths, along the $x$ or $y$ direction \cite{OVC2}.

Therefore at a very strong and a very weak coupling between two junctions the system behaves as effectively one-dimensional. Essential features of two dimensions are exhibited for an intermediate coupling, $\alpha\sim 1$. Below we study a formation of that regime starting from a region of small $\alpha$.
\subsection{Uncoupled junctions}
Let us consider first zero coupling between two junctions ($\alpha=0$). In this case Eqs.~(\ref{4}) and (\ref{6}) describe two independent particles in one-dimension potentials shown in Fig.~\ref{fig1}. When the semiclassical parameter $B$ is large the potential barriers in Fig.~\ref{fig1} are hardly transparent and a number of discrete levels in the wells is large.

Suppose tunneling to occur in the $x$ direction. We introduce the dimensionless energy $E$
\begin{equation}
\label{7d}
E_{0}=\frac{\hbar B}{t_{0}}E.
\end{equation}
The total particle energy $E$ in Fig.~\ref{fig1} is a sum of ones corresponding to motions in the $x$ and $y$ directions. A maximal tunneling probability of a particle, with a fixed total energy $E$, is realized when the energy of the $x$ motion has a maximal possible value. This situation is shown in Fig.~\ref{fig1}. In this case a motion in the direction perpendicular to tunneling (the $y$ direction) is not excited pertaining to a lowest level in $y$. In the classical language, the particle hits a border of the well, $V(x,y)=E$, with zero tangent velocity.

For comparison, in Fig.~\ref{fig2} the total energy $E$ is distributed in a way that the $y$ motion is excited. In the classical language, the particle hits a border of the well with a finite tangent velocity. This results in less probable tunneling in the $x$ direction since tunneling occurs from a lower level, $E-\delta E$, as shown in Fig.~\ref{fig2}. The
part $\delta E$ of the total energy relates to the tangent motion.
\begin{figure}
\includegraphics[width=5cm]{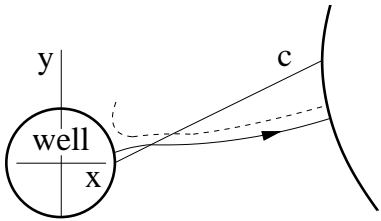}
\caption{\label{fig3}At $\delta E=0$ the main path (arrowed curve) connects two classically allowed domains, the potential well and the outer region. The thick curves relate to the
condition $V(x,y)=E$. A deviation from the main path (dashed curve) does not lead to the well. On the line $c$ the $y$ component of a force is zero, $\partial V/\partial y=0$.}
\end{figure}
\subsection{Weakly coupled junctions}
Suppose the junctions are weakly coupled so that $\alpha\ll~1$. In this case a motion in the total potential (\ref{6}) cannot be reduced to two independent ones as in Figs.~\ref{fig1} and \ref{fig2}. Now the entire system of levels in the total potential (\ref{6}) accumulates the levels of the $x$ and $y$ channels in Fig.~\ref{fig1}. If there are five discrete levels in each well, $V_{0}(x)$ and $V_{0}(y)$, then the potential $V(x,y)$ contains 25 levels.

Let us analyze specificity of two-dimensionality in tunneling. First, the coupling constant $\alpha$ is partly accounted for in the parameter $B\sim(1+\alpha)^{5/2}$ in Eq.~(\ref{4}). At small $\alpha$ this leads to a linear in $\alpha$ reduction of tunneling probability. The last term in the potential (\ref{6}) is proportional to $\alpha^{2}$ since $y\sim\alpha$. So we calculate an $\alpha^{2}$ correction to the tunneling probability. It is small at $\alpha\ll 1$ but it becomes significant at $\alpha\sim 1$.

Below in this subsection we propose some non-rigorous arguments which help to understand what happens under the barrier in two dimensions.

When a particle hits a border of the well with zero tangent momentum, $\delta E=0$, an underbarrier wave function is localized on the certain classical trajectory which is orthogonal to the
borders of classical regions \cite{COLEMAN1,COLEMAN2,SCHMID1,SCHMID2,MELN}. This trajectory is shown in Fig.~\ref{fig3} by the arrowed curve. The trajectory is driven by the $y$ force
directed away form the line $y=\alpha x/(1+\alpha)$ where it changes sign. That line is marked in Fig.~\ref{fig3} as $c$. Not far from the outer region the $x$ force, $\partial V/\partial x$, attracts the particle to the line $c$ as it is tilted.

A wave function is proportional to $\exp(iS/\hbar)$ where $S$ is a classical action \cite{LANDAU}. According to Maupertuis' principle \cite{LANDAU1}, along the classical trajectory one should put $S=S_{cl}$ where
\begin{figure}
\includegraphics[width=5cm]{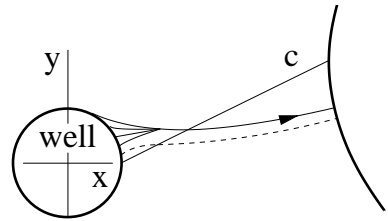}
\caption{\label{fig4}Underbarrier paths at $\delta E\neq 0$ connecting two classically allowed domains as in Fig.~\ref{fig3}. A finite tangent component results in the multi-path region
close to the well which goes over into the subsequent path indicated by the arrowed curve. The main path of Fig.~\ref{fig3} is shown by the dashed curve.}
\end{figure}
\begin{equation}
\label{7a}
S_{cl}=i\hbar B\int dl\sqrt{V(x,y)-E}
\end{equation}
and $dl$ is an element of the classical trajectory. This mechanism can be called main path tuneling since another path does not lead to the well in Fig.~\ref{fig3} deviating from it as the classical trajectory shown by the dashed curve \cite{SCHMID1,SCHMID2}. Eq.~(\ref{7a}) is generic with a conventional WKB result if to consider it along a curve but not a straight line.

A scenario changes when a particle hits the border of the well with non-zero tangent momentum, $\delta E\neq0$. Then an individual trajectory, related to the regime (\ref{7a}) and shown by the arrowed curve in Fig.~\ref{fig4}, is impossible at least in a vicinity of the well where a tangent component of the momentum is finite
\begin{equation}
\label{7b}
\frac{\partial S}{\partial y}\simeq\hbar B\sqrt{\delta E-V_{0}(y)}
\end{equation}
and the imaginary momentum in the $x$ direction is
\begin{equation}
\label{7c}
\frac{\partial iS}{\partial x}\simeq -\hbar B\sqrt{V_{0}(x)-E+\delta E}.
\end{equation}

In this case an attempt to adjust a classical trajectory to a tangent component at the well border results in a deviating curve similar to the dashed one in Fig.~\ref{3}. Therefore an underbarrier density cannot be localized on a particular classical trajectory along the entire underbarrier region. Close to the well that trajectory goes over into a wide set of paths as shown schematically in Fig.~\ref{fig4}. According to Feynman \cite{FEYN}, the state related to (\ref{7b}) and (\ref{7c}) is a superposition of a wide set of classical paths.

As follows from Eq.~(\ref{7c}), a finite $\delta E$ reduces the wave function in a vicinity of the well. In contrast, far from the well the wave function $\exp(iS_{cl}/\hbar)$ on the arrowed curve in Fig.~\ref{fig4} is larger compared to the dashed curve ($\delta E=0$) since the former is closer to the line $c$ where $V(x,y)$ is smaller. In other words, the arrowed curve in Fig.~\ref{fig4} is shifted to a more transparent part of the barrier.

When the energy $E$ is close to the bottom of the well the fraction of the trajectory above the curve $c$ in Fig.~\ref{3} is small because the well shrinks.

As follows from above, there are two opposite effects on tunneling when the total energy in the well is redistributed between a normal ($E-\delta E$) and a tangent ($\delta E$) motions.

(i) The reduction of tunneling is connected to a vicinity of well where the wave function is suppressed due to sinking of an energy level down to $E-\delta E$. As follows from Eq.~(\ref{7c}),
the exponential reduction of the particle density is $\exp(-c_{1}B\delta E)$ where $c_{1}$ is a positive parameter.

(ii) The enhancement of tunneling is due to a shift of the subsequent trajectory to a more transparent part of the barrier. A reduction of $V(x,y)$ in that region is proportional to
$\alpha xy$. Since a typical $x\sim 1$ and $y\sim\sqrt{\delta E}$, the exponential enhancement of the particle density, according to Eq.~(\ref{7a}), can be estimated as $\exp(2c_{2}B\alpha\sqrt{\delta E)}$ where $c_{2}$ is a positive parameter.

A total effect on the tunneling probability $\Gamma(\delta E)$ is defined by a product of the two exponents
\begin{equation}
\label{7e}
\Gamma(\delta E)\sim\Gamma(0)\exp(-c_{1}B\delta E)\exp(2c_{2}B\alpha\sqrt{\delta E}).
\end{equation}
At $\delta E\sim\alpha^{2}$ the expression (\ref{7e}) reaches a maximum, $\Gamma(0)\exp(B\alpha^{2}c^{2}_{2}/c_{1})$, which manifests an exponential enhancement of tunneling due to a finite $\delta E$. The parameters $c_{1}$ and $c_{2}$ are approximately of the order of unity. Whereas $\alpha$ is small the semiclassical combination $B\alpha^{2}$ is large.

It is amazing, that the above conclusions, drawn on the basis of general arguments, are confirmed (excepting some details) by exact calculation in Sec.~\ref{ham-jac} where the values of $c_{1}$ and $c_{2}$ are specified.
\subsection{Analogy with photon-assisted tunneling}
Multi-path tunneling through a two-dimensional static barrier reminds photon-assisted tunneling across a nonstationary one-dimensional barrier \cite{MELN1}. The latter also consists of two parts. The first one is an absorption of quanta with an exponentially small probability analogous to the first exponent in Eq.~(\ref{7e}). The second one is tunneling in a more transparent part of the barrier (a higher energy) with an enhanced probability corresponding to the second exponent in Eq.~(\ref{7e}). A total probability is also determined by a maximum of the product.

In the both cases a particle finds a more transparent part of a barrier being initially pushed either by a tangent motion (at the end of the multi-path region in Fig.~\ref{fig4}) or by quanta absorption.
\section{UNDERBARRIER WAVE FUNCTION}
\label{ham-jac}
To quantitatively study the problem of two-dimensional tunneling one should solve the Schr\"{o}dinger equation with the exact potential (\ref{6}). Since the potential barrier is almost classical one can apply a semiclassical method. With an exponential accuracy the wave function is
\begin{equation}
\label{8}
\psi\sim\exp(iB\sigma\sqrt{2}),
\end{equation}
where the exponent is a large classical action measured in the units of Planck's constant. As follows from the form (\ref{4}), $\sigma(x,y)$ satisfies the Hamilton-Jacobi equation \cite{LANDAU}
\begin{equation}
\label{9}
\left(\frac{\partial\sigma}{\partial x}\right)^{2}+\left(\frac{\partial\sigma}{\partial y}\right)^{2}+V_{0}(x)+V_{0}(y)-f(x)y=E,
\end{equation}
where $f(x)=2\alpha x/(1+\alpha)$. When $\alpha=0$ the variables in Eq.~(\ref{9}) are separated and a solution can be easily obtained. In our case there a substantial cross-term $f(x)y$ in  Eq.~(\ref{9}) which mixes the modes.

We consider a small coupling $\alpha$ when a deviation of the variable $y$ from the tunneling path is small. For convenience, below a model coupling between the variables $x$ and $y$
is introduced. Namely, instead of the linear in $x$ function we use
\begin{equation}
\label{12}
f(x)=
\begin{cases}
0,& x<x_{0}\\
2\alpha,& x_{0}<x,
\end{cases}
\end{equation}
where $x_{0}$ is chosen between $a_{2}$ and $a_{1}$ in Fig.~\ref{fig1}. The choice of the form (\ref{12}) does not contradict to main arguments of Sec.~\ref{gen}.
\subsection{Hamilton-Jacobi approach}
At a small coupling, $\alpha\ll 1$, we consider a transition through the barrier along the direction $x$. The transition occurs with a small deviation of $y$ from zero position. In this
case one can put $V_{0}(y)=y^{2}$ in Eq.~(\ref{9}). The classically allowed motion in the well $V_{0}(x)+y^{2}<E$ is described by a real action $\sigma(x,y)$. The particle 
probes the potential in the classically allowed region where motions in $x$ and $y$ directions are independent. Therefore the level quantization is determined by two one-dimensional 
wells and the action is a sum of two one-dimensional parts defined by a solution of Eq.~(\ref{9}) with the condition (\ref{12}) at $x<x_{0}$. A continuation of this action from the well to under the barrier at $x<x_{0}$ results in
\begin{figure}
\includegraphics[width=5cm]{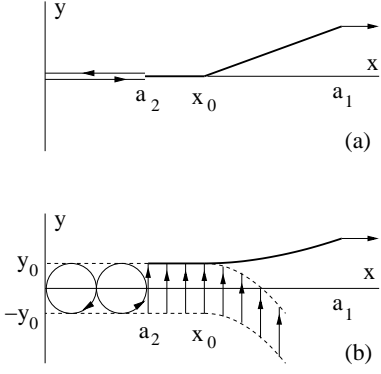}
\caption{\label{fig5}Real momenta are indicated by arrows. A maximum of the particle density under the barrier is reached on the thick curves $y(x)$. (a) Tunneling from a state with zero
tangent momentum, $\delta E=0$. (b) Tunneling from a state with non-zero tangent momentum, $\delta E\neq 0$ when $y_{0}=\sqrt{\delta E}$ is finite. Real momenta at the well ($x<a_{2}$) are
shown on a classical trajectory only.}
\end{figure}
\begin{equation}
\label{13}
\sigma(x,y)=i\int^{x}_{x_{0}}dx_{1}\sqrt{V_{0}(x_{1})-E+\delta E}+\int^{y}_{0}dy_{1}\sqrt{\delta E-y^{2}_{1}}
\end{equation}
The part $\delta E$ of the total energy $E$ relates to the tangent (perpendicular to the tuneling direction) motion. The fraction $\delta E$ is determined by a state in the well from which
tunneling occurs. At a large $B$ the energies $E$ and $\delta E$ are almost continuous with a discreetness $1/B$.

A solution at $x_{0}<x$ can be found by the method of variation of constants \cite{LANDAU1}. It has the form
\begin{eqnarray}
\label{14}
&&\sigma(x,y)=i\int^{x}_{x_{0}}dx_{1}\sqrt{V_{0}(x_{1})-E-\alpha^{2}+\varepsilon(x,y)}\\
\nonumber
&&+\int^{y}_{\alpha}dy_{1}\sqrt{\varepsilon(x,y)-(y_{1}-\alpha)^{2}}+iF[\varepsilon(x,y)],\hspace{0.2cm}x_{0}<x.
\end{eqnarray}
For a given function $F(\varepsilon)$ the function $\varepsilon(x,y)$ is determined by the condition $\partial\sigma/\partial\varepsilon=0$
\begin{eqnarray}
\label{15}
&&2i\frac{\partial F(\varepsilon)}{\partial\varepsilon}=
i\int^{x}_{x_{0}}\frac{dx_{1}}{\sqrt{V_{0}(x_{1})-E-\alpha^{2}+\varepsilon}}\\
\nonumber
&&+\int^{y}_{\alpha}\frac{dy_{1}}{\sqrt{\varepsilon-(y_{1}-\alpha)^{2}}},
\end{eqnarray}
which is independence of $\sigma$ on ``constant'' $\varepsilon(x,y)$. Now derivatives of the action have the simple forms
\begin{eqnarray}
\label{16}
&&\frac{\partial\sigma(x,y)}{\partial x}=i\sqrt{V_{0}(x)-E-\alpha^{2}+\varepsilon(x,y)}\\
\nonumber
&&\frac{\partial\sigma(x,y)}{\partial y}=\sqrt{\varepsilon(x,y)-(y-\alpha)^{2}}.
\end{eqnarray}
According to matching of $\partial\sigma(x_{0},y)/\partial y$ given by (\ref{13}) and (\ref{14}), $\varepsilon(x_{0},y)=\delta E+\alpha^{2}-2\alpha y$. This can be used to determined the function $F(\varepsilon)$ if to express $y$ through $\varepsilon(x_{0},y)$ and to insert that into Eq.~(\ref{15}). As a result, the function $\varepsilon(x,y)$ is defined by the condition
\begin{equation}
\label{17}
i\int^{x}_{x_{0}}\frac{dx_{1}}{\sqrt{V_{0}(x_{1})-E-\alpha^{2}+\varepsilon}}=
\int^{(\delta E-\alpha^{2}-\varepsilon)/2\alpha}_{y-\alpha}\frac{dz}{\sqrt{\varepsilon-z^{2}}}
\end{equation}
\subsection{Probability of tunneling}
Not very close to the classical exit point $a_{1}$, determined by the condition $V_{0}(x)\simeq E$, one can write the left hand side of Eq.~(\ref{17}) as
$i\tau(x)$ where 
\begin{equation}
\label{18}
\tau(x)=\int^{x}_{x_{0}}\frac{dx_{1}}{\sqrt{V_{0}(x_{1})-E}}.
\end{equation}
Now a solution of Eq.~(\ref{17}) can be easily obtained. At $x_{0}<x$
\begin{equation}
\label{19}
\sqrt{(y-\alpha)^{2}-\varepsilon}=\sqrt{(y+2\alpha\sinh^{2}\tau/2)^{2}-\delta E}-\alpha\sinh\tau.
\end{equation}
There is a remarkable underbarrier path, $y(x)$, where $\partial\sigma(x,y)/\partial y=0$. According to Eq.~(\ref{16}), at $x_{0}<x$
\begin{equation}
\label{20}
y(x)=\sqrt{\alpha^{2}\sinh^{2}\tau(x)+\delta E}-\alpha[\cosh\tau(x)-1].
\end{equation}
The path $y(x)$ is shown in Fig.~\ref{fig5} by thick curves.

The region $x<a_{2}$ pertains to the classical motion in the well. At $a_{2}<x<x_{0}$ a particle density decays away from the region $-\delta E<y<\delta E$. At $x_{0}<x$ the underbarrier density reaches a maximum (with respect to $y$) on the curve $y(x)$. At a finite $\delta E$ there are real momenta under the barrier indicated by arrows in Fig.~\ref{fig5}(b). At $x_{0}<x$ at the region with real momenta the wave function is exponentially smaller than on the curve $y(x)$. So at $x_{0}<x$ one can account for the path $y(x)$ only. It is generic with the arrowed curve in Fig.~\ref{fig4}. At $a_{1}<x$ the particle escapes from under the barrier.

Below it is convenient to introduce
\begin{equation}
\label{21}
\mu(x)=\varepsilon[x,y(x)].
\end{equation}
As follows from Eqs.~(\ref{19}) and (\ref{20}),
\begin{equation}
\label{22}
\sqrt{\mu(x)}=\alpha\cosh\tau(x)-\sqrt{\alpha^{2}\sinh^{2}\tau(x)+\delta E}.
\end{equation}

We define probability of tunneling with a fixed energy $E$ and an exchange $\delta E$ as
\begin{eqnarray}
\label{23}
\nonumber
&&\Gamma(\delta E)=\bigg|\frac{\psi[a_{1},y(a_{1})]}{\psi(a_{2},0)}\bigg|^{2}\\
&&\sim\exp\left(2iB\sqrt{2}\int^{a_{1}}_{a_{2}}\frac{\partial\sigma[x,y(x)]}{\partial x}dx\right).
\end{eqnarray}
The definition (\ref{23}) is analogous to Eq.~(\ref{7e}). It is usefull to consider the probability of tunneling $\Gamma(0)$ at $\delta E=0$. To obtain it one has to expand the root in the first equation (\ref{16}) with respect to $\varepsilon-\alpha^{2}$ and to substitute into Eq.~(\ref{23}). With the use of Eqs.~(\ref{21}) and (\ref{22}) we obtain
\begin{equation}
\label{24}
\Gamma(0)=\Gamma_{WKB}\exp\left\{\frac{B\alpha^{2}}{\sqrt{2}}\left[2\tau(a_{1})-1+e^{-2\tau(a_{1})}\right]\right\},
\end{equation}
where
\begin{equation}
\label{25}
\Gamma_{WKB}\sim\exp\left[-2B\sqrt{2}\int^{a_{1}}_{a_{2}}dx\sqrt{V_{0}(x)-E}\right]
\end{equation}
is a one-dimensional WKB like expression.
\begin{figure}
\includegraphics[width=4.5cm]{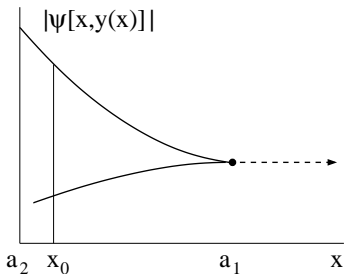}
\caption{\label{fig6}Modulus of the wave function at $\delta E=\alpha^{2}$. In this case $y(x)=\alpha$ at all $a_{2}<x<a_{1}$.}
\end{figure}
\subsection{Tunneling from a state with non-zero tangent momentum, $\delta E\neq 0$}
We consider weak tangent momenta related to the condition $\delta E\sim\alpha^{2}$.

According to Eqs.~(\ref{23}), (\ref{16}), and (\ref{22}), at $\delta E\ll\alpha^{2}$ the tunneling probability is
\begin{equation}
\label{29}
\Gamma(\delta E)=\Gamma(0)\exp\left(-B\delta E\sqrt{2}|\tau(a_{2})|+\frac{B\delta E}{\sqrt{2}}\ln\frac{\alpha^{2}}{\delta E}\right).
\end{equation}
The first term in the exponent comes from a conventional WKB reduction of tunneling probability at $x<x_{0}$ when the energy is reduced by $\delta E$. The second term in the exponent is due to
the region $x_{0}<x$ where the trajectory goes in a more transparent part of the barrier. The second term dominates at $\delta E\ll\alpha^{2}$ and increases the tunneling probability.

As follows from Eq.~(\ref{22}), at $\delta E$ close to $\alpha^{2}$, $\mu(x)=(\alpha-\sqrt{\delta E})^{2}/\cosh^{2}\tau(x)$. This means that the increase of the tunneling probability,
associated with the region $x_{0}<x$, is more effective at $\delta E=\alpha^{2}$. According to Sec.~\ref{gen}, when $E$ is not far from the well bottom the crossover between two regimes is shifted toward the point $x=a_{2}$. For this reason, below we choose the parameter $x_{0}$ in Eq.~(\ref{12}) to be close to $a_{2}$ for simplicity. In this case, at $\delta E$ close to $\alpha^{2}$, the tunneling probability is
\begin{eqnarray}
\nonumber
&&\Gamma(\delta E)=\Gamma(0)\exp\bigg\{\frac{B\alpha^{2}}{\sqrt{2}}\Big[1-e^{-2\tau(a_{1})}\\
\label{30}
&&-\frac{2}{\alpha^{2}}(\alpha-\sqrt{\delta E})^{2}\tanh\tau(a_{1})\Big]\bigg\}.
\end{eqnarray}

There is another expression of the probability which follows from Eqs.~(\ref{24}) and (\ref{30})
\begin{eqnarray}
\nonumber
&&\Gamma(\delta E)=\Gamma_{WKB}\exp\bigg\{B\alpha^{2}\sqrt{2}\bigg[\tau(a_{1})\\
\label{30a}
&&-\frac{(\alpha-\sqrt{\delta E})^{2}}{\alpha^{2}}\tanh\tau(a_{1})\bigg]\bigg\}.
\end{eqnarray}
The Gaussian form (\ref{30a}) is analogous to Eq.~(\ref{7e}). The tunneling probability reaches a maximum at $\delta E=\alpha^{2}$ as it has been predicted in Sec.~\ref{gen}. The modulus of the wave function is plotted in Fig.~\ref{fig6} where $y(x)=\alpha$ for all underbarrier track, $a_{2}<x<a_{1}$. There are two branches under the barrier and an outgoing wave after the exit indicated by the dashed line in Fig.~\ref{fig6}.
\subsection{Temperature dependence of tunneling probability}
At a fixed particle energy $E$ a tunneling probability is determined by Eqs.~(\ref{30a}) and (\ref{25}) with $\delta E=\alpha^{2}$. At a fixed temperature all energies contribute to tunneling with Gibbs factors $\exp(-E_{0}/T)$. In the semiclassical limit, $1\ll B$, there are many levels in the well. Therefore one can treat $E$ as a continuous variable and to optimize the probability $\Gamma(\delta E)\exp(-E_{0}/T))$ with respect to $E$ accounting for Eq.~(\ref{7d}). This procedure gives the certain optimal energy $E_{T}$ from which tuneling occurs
\cite{LEGGETT,LEGGETT1,COLEMAN1,COLEMAN2,SCHMID1,SCHMID2}. If to omit the term proportional to $\alpha^{2}$ in $\Gamma(\delta E)$ (\ref{30a}) an optimal energy $E_{T}$ is determind by
\begin{equation}
\label{30b}
\tau(a_{1})=\frac{\hbar}{t_{0}T\sqrt{2}}.
\end{equation}
Imaginary time $\tau(a_{1})$ is given by Eq.~(\ref{18}) where one should put now $x_{0}=a_{2}$. It follows that at $T=0$ the optimal energy is $E_{T}=0$. $E_{T}$ at the barrier top in Fig.~\ref{fig2} corresponds, at small $\alpha$, to the critical temperature $T_{0}=\hbar/\pi t_{0}\sqrt{2}$ when the decay occurs solely due to thermal activation.

Now one can express the tuneling probability $\Gamma_{T}$ at a fixed temperature in the form
\begin{equation}
\label{30c}
\Gamma_{T}=\Gamma^{(0)}_{T}\exp\left\{\frac{B\alpha^{2}}{\sqrt{2}}\left[1-\exp\left(-\frac{2\pi T_{0}}{T}\right)\right]\right\},
\end{equation}
where the exponential coinsides with one of Eq.~(\ref{30}) if to insert there $\delta E=\alpha^{2}$ and the expression (\ref{30b}). $\Gamma^{(0)}_{T}$ is a probability of a conventional tunneling when a tangent momentum in the well is zero. As follows from the expression (\ref{30c}), $\Gamma^{(0)}_{T}<\Gamma_{T}$. This means that tunneling occurs from states with a finite tangent momentum.

Since energy $E$ has to be larger than $\delta E(=\alpha^{2})$ temperature cannot be too low, $1/\ln(1/\alpha)<T/T_{c}<1$.
\subsection{Dependence of tunneling probability on current}
Let us consider first an almost continuous distribution of levels in the well ($1\ll B$) as in subsection D.

\begin{figure}
\includegraphics[width=4.5cm]{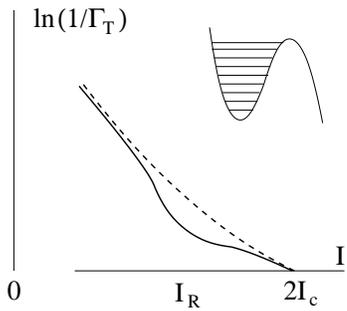}
\caption{\label{fig7}Dependence of tunneling probability at a fixed temperature $\Gamma_{T}$ on current in the case of almost continuous distribution of levels in the well (see insert). The dashed curve corresponds to the conventional probability $\ln(1/\Gamma^{(0)}_{T})$.}
\end{figure}

The coupling constant $\alpha$ is of the order of unity at $I=I_{R}$ where
\begin{equation}
\label{31}
\left(1-\frac{I_{R}}{2I_{c}}\right)=\frac{1}{\beta^{2}}.
\end{equation}
Since currents are close to the critical value $2I_{c}$ the parameter $\beta$ is supposed to be large. A dependence of $\Gamma_{T}$ on current is schematically shown in Fig.~\ref{fig7}.
The limit of a small $\alpha$, considered above and defined by Eq.~(\ref{30c}), corresponds to the left part of the curves in Fig.~\ref{fig7}. Parts of the curves in Fig.~\ref{fig7} closer to $2I_{c}$ pertain to a large $\alpha$. In this case tunneling occurs symmetrically, $x=y$, since a deviation from this path costs a lot of energy \cite{OVC2}.

The dashed curve in Fig.~\ref{fig7} is $\Gamma^{(0)}_{T}$. It is very easy to numerically calculate $\Gamma^{(0)}_{T}$ using a standard technique \cite{OVC2} when a two-dimensional trajectory is normal to classically allowed regions \cite{LEGGETT,LEGGETT1,BLAT,COLEMAN1,COLEMAN2,SCHMID1,SCHMID2}. $\Gamma^{(0)}_{T}$ depends on parameters (coupling strength, asymmetry of junctions, etc.) of a SQUID fabricated for measurements. For comparison with experiments one has to know those values exactly. For this reason, we leave $\Gamma^{(0)}_{T}$ as a schematic dashed curve in Fig.~\ref{fig7} planning its exact calculation in the nearest future for particular parameters of a SQUID with an intermediate effective coupling.

The curves in Fig.~\ref{fig7} are taken at a fixed temperature which exceeds $T_{0}=[2(1-I/I_{c})]^{1/4}\hbar\omega/2\pi$ sufficiently close to $2I_{c}$ (large $\alpha$) \cite{OVC2}. In other words, in a close vicinity of $2I_{c}$ the decay is due to a thermal activation. We will consider details elsewhere.

When the semiclassical parameter $B$ is not too large a tunneling probability still remains exponentially small but a level distribution in the well becomes substantially discrete. This is a typical experimental situation \cite{CLARKE3,CLARKE4,UST1,UST2}. The underbarrier process sets the certain energy $\delta E=\alpha^{2}$ which is given up to the tangent motion. According to Eq.~(\ref{30}), the tunneling probability reaches a maximal value under that condition. The energy exchange is similar to one illustarted in Fig.~\ref{fig2}.

On the other hand, the optimal $\delta E$ may not be exactly fitted by discrete energy levels in the well which can be outside the Gaussian distribution (\ref{30}). This contrasts to a continuous distribution of levels at a large $B$ when a proper level always exists. A mismatch between $\delta E$ and the level structure results in reduction of the tunneling
probability.
\begin{figure}
\includegraphics[width=4.5cm]{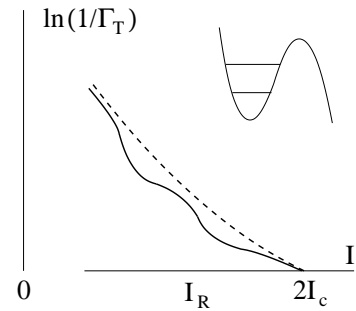}
\caption{\label{fig8}Dependence of tunneling probability at a fixed temperature $\Gamma_{T}$ on current in the case of discrete levels in the well (see insert). The dashed curve is the same as in Fig.~\ref{fig7}.}
\end{figure}

For this reason, one can expect a "resonance" due to coincidence of $\delta E(=\alpha^{2})$ with a distance between discrete levels. There are no sharp peaks but rather a shallow wavy structure 
in $\Gamma_{T}$ illustrated in Fig.~\ref{fig8}. Each minimum results from the "resonance" and is smeared out at the interval of currents
\begin{equation}
\label{32}
\frac{\delta I}{2I_{c}-I_{R}}\sim\frac{1}{\sqrt{B}},
\end{equation}
which follows from the Gaussian distribution in Eq.~(\ref{30}). Under the condition (\ref{31}) a depth of each minimum in Fig.~\ref{fig8} is of the order of one. In Fig.~\ref{fig8} only two minima are shown.

Strictly speaking, the semiclassical theory is not applicable to a case of moderate $B$ and the above arguments are heuristic.
\section{DISCUSSIONS}
\label{disc}
Quantum tunneling across a one-dimensional static potential barrier is described by WKB theory. Tunneling through a multi-dimensional barrier is well studied 
\cite{COLEMAN1,COLEMAN2,SCHMID1,SCHMID2}. Accordingly, the main contribution to a tuneling probability comes from the extreme path linking two classically allowed regions. The path is a classical trajectory with real coordinates which can be parametrized by imaginary time. The underbarrier trajectory is a solution of Newton's equation in imaginary time. The trajectory is given rise by a particle hitting normally (with zero tangent momentum) a border of the classically allowed region. In terms of discrete levels in the well tunneling occurs from a state with the analogous property. Under the barrier the probability density reaches a maximum at each point of the trajectory along the orthogonal direction with respect to it. Therefore around the trajectory, which plays a role of a saddle point, quantum fluctuations are weak. The wave function, tracked along that trajectory under the barrier, exhibits an exponential decay generic with WKB behavior. The above mechanism, which can be called main-path tunneling, was explored for two-dimensional tunneling in a SQUID in Ref.~\cite{OVC2}.

However, in some cases tunneling through multi-dimensional barriers occurs according to a different scenario which is far from being similar to WKB. An example of such situation in a SQUID is investigated in the paper.

When a state before tunneling has a tangent momentum with respect to a border of the well there are no extreme points on it since the derivative of the wave function along the border is finite. This means that a tunneling probability is no more determined by the main underbarrier path but comes from a wide set of paths. Traditionally, a decay of a state with a
tangent momentum is not considered since it does not correspond to a saddle point and, hence, the net contribution is expected to be averaged down to a small value due to mutual interference
of trajectories.

It is shown in the paper by general arguments and exact calculations that a non-zero tangent momentum does not result in reduction of tunneling probability. Moreover, a related multi-path mechanism can exponentially enhance barrier penetration in a SQUID. In contrast to main-path mechanism, the multi-path tunneling cannot be calculated using a classical trajectory in imaginary time. Such a trajectory, connecting two physical (real) points, does not exist in that case.

A possible way of interpretation of multi-path tunneling is proposed in Sec.~\ref{gen}. In a vicinity of an enter under the barrier it is locally less transparent. A wide distribution of paths in that region goes over into the single path which proceeds up to an end of the barrier. That path lies in a more transparent part of the barrier resulting in enhancement of tunneling probability.

Multi-path tunneling through a two-dimensional static barrier reminds photon-assisted tunneling across a nonstationary one-dimensional barrier. In the both cases a particle finds a more transparent part of a barrier being initially pushed either by a tangent motion or by quanta absorption.

The total energy $E$ of a state in the well is distributed as $E-\delta E$ for a motion in the tunneling direction and as $\delta E$ for tangent one. For a small coupling constant $\alpha$
the maximal tunneling probability corresponds to $\delta E\sim\alpha^{2}$.

An analysis of experimental data allows to define a contribution of multi-path effects by separation of the conventional probability $\Gamma^{(0)}_{T}$ from a total result. $\Gamma^{(0)}_{T}$ depends on parameters (coupling strength, asymmetry of junctions, etc.) of a particular SQUID fabricated for measurements.

A SQUID should be with a large parameter $\beta$. The calculations in the paper are done for a symmetric SQUID in zero magnetic field but this is not a principle restriction for observation of multi-path tunneling. A role of dissipation is to be studied.

We see that in a multi-dimensional system (a SQUID is a two-dimensional example) tunneling mechanism can be different than one traditionally considered on the basis of a main
trajectory. There are two necessary condition for unusual tunneling. First, tunneling should not start from a ground state, otherwise an energy exchange is impossible. Second, coordinates
should not be almost separated when tunneling occurs practically along one of them (the limits of small and large coupling in a SQUID).

One should expect a modified quantum tunneling at a finite temperature across a non-one-dimensional barrier when coordinates are not separated. In this case the conventional mechanism of a periodic trajectory in imaginary time with the period $\hbar/T$ is substituted by multi-path.

\section{CONCLUSION}
Traditionally quantum tunneling in a static SQUID is studied on the basis of a classical trajectory in imaginary time under a two-dimensional potential barrier. The trajectory connects
a potential well and an outer region crossing their borders in perpendicular directions. In contrast to that main-path mechanism, a wide set of trajectories with components tangent to the border of the well can constitute an alternative mechanism of multi-path tunneling. The phenomenon is essentially non-one-dimensional. Continuously distributed paths under the barrier result in enhancement of tunneling probability. A type of tunneling mechanism (main-path or multi-path) depends on character of a state in the potential well prior to tunneling.

\acknowledgments
I thank A. V. Ustinov for valuable discussions.

\end{document}